\documentclass[pdflatex,sn-mathphys-num]{sn-jnl}

\usepackage{graphicx}%
\usepackage{multirow}%
\usepackage{amsmath,amssymb,amsfonts}%
\usepackage{amsthm}%
\usepackage{mathrsfs}%
\usepackage[title]{appendix}%
\usepackage{xcolor}%
\usepackage{textcomp}%
\usepackage{manyfoot}%
\usepackage{booktabs}%
\usepackage{algorithm}%
\usepackage{algorithmicx}%
\usepackage{algpseudocode}%
\usepackage{listings}%
\usepackage{float}
\usepackage{comment}
\usepackage{lineno}

\theoremstyle{thmstyleone}%
\theoremstyle{thmstyletwo}%

\theoremstyle{thmstylethree}%

\begin{document}
\title[Article Title]{A cavity-less architecture for high-power integrated frequency combs}

\author[1]{\fnm{Mrinmoy} \sur{Roy}}
\equalcont{These authors contributed equally to this work.}
\author[1]{\fnm{Joshua A.} \sur{Palacios}}
\equalcont{These authors contributed equally to this work.}
\author[1]{\fnm{Shuva} \sur{Roy}}
\author*[1]{\fnm{Darren D.} \sur{Hudson}}\email{darren.hudson@ucf.edu}
\author*[1]{\fnm{Andrea} \sur{Blanco-Redondo}}\email{andrea@creol.ucf.edu}

\affil[1]{\orgdiv{CREOL, The College of Optics and Photonics}, \orgname{University of Central Florida}, \orgaddress{\street{4304 Scorpius St.}, \city{Orlando}, \postcode{32816}, \state{FL}, \country{USA}}}


\abstract{Photonic chip-based frequency combs have emerged as a transformative platform, enabling compact, scalable, and high-performance multiwavelength sources with far-reaching impact across science and technology. Most commonly, these sources leverage the cavity enhancement of the nonlinearities to produce a spectrum of equidistant frequency lines via cascaded four-wave mixing in high-quality microresonators pumped with a continuous wave tone. While the presence of the resonator inherently enables low-power threshold operation, it also brings intrinsic limitations in efficiency, tunability, and power per line. Here, we propose and demonstrate a cavity-less approach for the generation of optical frequency combs on-chip, which relies on non-degenerate cascaded four wave mixing in dispersion engineered integrated photonic waveguides. The results presented here enable previously inaccessible regimes of pump-to-comb conversion efficiency, wide-range continuous line-spacing tunability, and power per line for coherent comb states. This work opens new research opportunities in nonlinear integrated photonics and pathways toward high-capacity optical interconnects, scalable photonic AI accelerators, and other power-constrained integrated systems.}

\maketitle

\section{Introduction}\label{sec1}

Optical frequency combs $-$ phase-coherent spectra of evenly spaced optical frequencies $-$ have revolutionized precision metrology since their inception \cite{Jones2000, udem2002optical}. Although initially associated with self-referenced combs for timekeeping, the term now broadly describes light sources whose spectra consist of discrete, uniformly spaced frequency components \cite{diddams2020optical}. Microresonator frequency combs, also known as microcombs \cite{del2007optical, kippenberg2011microresonator, gaeta2019photonic, chang2022integrated}, have enabled the miniaturization of this technology making it attractive for a wide range of applications where size, power, repetition rate, and integrability are the enabling factors \cite{obrzud2019microphotonic, marin2017microresonator, suh2016microresonator, suh2018soliton}, and unveiling a variety of novel nonlinear optical phenomena \cite{cole2017soliton, lucas2017breathing, yi2017single}. In their most common form, microcombs rely on pumping a high-quality (high-$Q$) microresonator with a continuous-wave (CW) laser, where Kerr $\chi ^{(3)}$ nonlinearity converts the pump into a set of frequency lines whose position is defined by the resonator’s free spectral range (FSR) via cascaded four-wave mixing (CFWM). The use of high-Q microresonators enables operation at very low pump-power thresholds thanks to the cavity's nonlinear enhancement \cite{del2007optical, PhysRevA.71.033804, ji2017ultra}.  However, the presence of the cavity also entails some limitations. First, the comb line spacing is determined by the FSR of the cavity and therefore fixed by the geometry at fabrication. Several schemes have been proposed to tune this line spacing discretely, in multiples of the cavity's FSR, including pump tuning \cite{raja2019electrically} and dual-pump schemes \cite{wang2016dual}. Continuous tunability has also been demonstrated in microcombs using thermal \cite{xue2016thermal} and electro-optic modulation \cite{del2012hybrid, he2023high}, albeit with a line-spacing tuning range below 100 GHz \cite{Shu2023}. A second limitation, specific to coherent micorcombs, is that their inherent pump to comb conversion efficiency is linked to the pump-resonance phase detuning, which generally leads to low efficiencies for coherent comb states such as solitons \cite{Xue2017LPR}. The conversion efficiency of coherent combs can be improved by using coupled ring resonator schemes or interferometric back-coupling, with record efficiencies of $\approx50\%$ reported for Kerr microcombs \cite{helgason2023surpassing, boggio2022efficient} and $\approx30\%$ for electro-optic combs \cite{hu2022high}.  Finally, attaining high on-chip power in microcombs and a sufficiently high amount of lines with power levels exceeding 100$\mu$W $-$ deemed as the minimum usable power level for applications such as dense wavelength division multiplexing (DWDM) $-$ has been a challenge, especially for high-coherence comb states \cite{gil2025high}.

As an alternative to the use of high-Q cavities with fixed FSR, here, we propose and demonstrate a new approach to on-chip comb generation based on non-degenerate CFWM (ND-CFWM) \cite{Crespo:00, McKinstrie2002, mckinstrie2006four} in dispersion-engineered integrated photonic waveguides. Drawing inspiration from earlier demonstrations of high-power broadband ND-CFWM spectra in fiber platforms \cite{winters2023octave,Thapa:24,Thapa2026}, we use a dual-oscillator power amplifier (DOPA) as our pump seed, which delivers quasi-CW tones with significantly increased peak power at a fixed average power. At its optimal operation point, our cavity-less comb produces 283 lines and spans 623 nm, with 62 of those lines carrying more than 100 $\mu$W, 29 lines $>$1 mW, and 3 lines $>$ 10 mW. Our comb also showcases inherently high pump-to-comb efficiency (79$\%$), continuous line-spacing tunability capability over a wide range (656 GHz), and a high degree of coherence, producing transform limited ultrafast soliton pulses. These features highlight this cavity-less approach to on-chip comb generation as an attractive technology for power-hungry applications demanding high efficiency and tunability, such as coherent optical communications, AI photonic accelerators, and LIDAR sensing. 

\section{Results}\label{sec2}
\subsection{Cavity-less comb generation}
\begin{figure}[H]
    \centering
    \includegraphics[width=\textwidth]{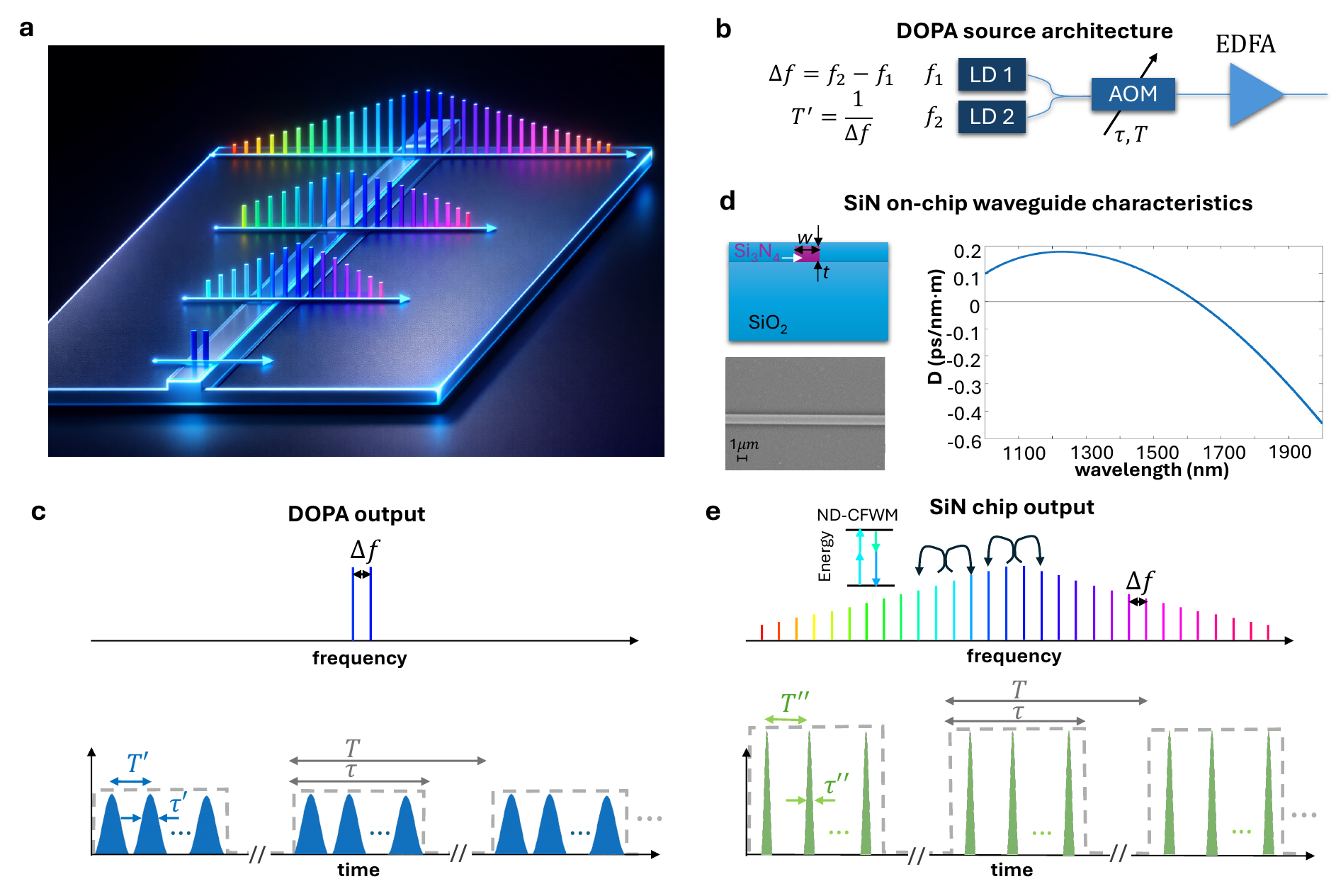}
    \caption{\textbf{Cavity-less comb generation main concepts. a}, Artist representation of the on-chip comb generation via non-degenerate four-wave mixing starting from two initial lines in a silicon nitride waveguide. \textbf{b}, Schematic of the dual-oscillator power amplifier (DOPA) seed comprising two tunable laser diodes (LD) combined and fed through an acousto-optic modulator (AOM), and through an erbium dooped fiber amplifier (EDFA). \textbf{c}, Conceptual representation of the frequency (top) and time (bottom) output of the DOPA source. \textbf{d}, Geometry of the silicon nitride waveguide used in the experiments, with width $w=940$ nm, thickness $t=812$ nm, and length of 5 mm (top left); Scanning electron micrograph (SEM) of the waveguide (bottom left); Simulated dispersion of the waveguide (right). \textbf{e}, Conceptual representation of the frequency (top) and time (bottom) output of the chip.}
    \label{fig:concept}
\end{figure}

The cornerstone of our comb approach is ND-CFWM in a dispersion-engineered integrated photonic waveguide (Fig. \ref{fig:concept}a). To initiate the ND-CFWM process, in the absence of a resonant structure, we use two laser tones at frequencies $f_1$ and $f_2$, respectively, whose separation $\Delta f = f_2-f_1$ will eventually determine the line-spacing of the comb. As the two tones propagate through a silicon nitride (SiN) waveguide with appropriate dispersion at the center frequency $f_0 = (f_1+f_2)/2$ new pairs of frequencies start to appear: first, we see the appearance of two new lines at $f_1-\Delta f$ and  $f_2+\Delta f$, and subsequently, through consecutive instances of CFWM an increasing number of lines at frequencies $f_1-n\Delta f$ and $f_2+n\Delta f$ arise, where $n\geq2$. 

Although, fundamentally, this scheme would work with two initial continuous wave (CW) frequency tones, in practice it is hard to achieve the necessary peak powers to achieve broad high-power per line frequency combs relying solely in CW pumping. To circumvent this limitation, our laser seed is a dual-oscillator power amplifier (DOPA) \cite{Thapa:24,Thapa2026} (Fig. \ref{fig:concept}b). The DOPA relies on two combined tunable CW laser diodes at frequencies $f_1$ and $f_2$ modulated by an acousto-optic modulator (AOM) and amplified by an erbium doped fiber amplifier (EDFA). The AOM carves long optical pulses of duration $\tau$ $-$ on the order of tens of nanoseconds $-$ at a repetition rate $1/T$ into the CW signal to avoid gain saturation in the subsequent amplifier and thus achieve higher peak power going into the integrated waveguide. In the frequency domain, the laser seed consists of two narrow linewidth frequency lines spaced by $\Delta f$ (Fig. \ref{fig:concept}c). The linewidth of these lines scales with $\propto\tau ^{-1} $ and can therefore be tuned by adjusting $\tau$. In the time domain, the DOPA outputs a beat pattern between the two frequencies, where each modulation peak has a width $\tau '$  and the repetition period is $T'=1/\Delta f$. 
This pattern is repeated in bursts of $\tau$ duration with a period $T$, yielding a peak power that is higher than the average power by a factor of $T/\tau$ which leads to a nonlinear enhancement analogous to that provided by the cavity in a traditional microcomb.



This laser seed is coupled to a dispersion-engineered SiN integrated photonic waveguide. The optimal dispersion profile for broadband high-power comb generation in the waveguide $-$ determined using nonlinear Schrödinger equation (NLSE) simulations with a two-tone excitation $-$ was found to be moderate anomalous dispersion at $f_0$ surrounded by two zero-dispersion points. To achieve this profile we performed dispersion-engineering on thick SiN waveguides $-$ where the thickness is fixed by available fabrication processes to $t\sim 800$ nm $-$ optimizing the waveguide width using an eigenmode solver (Lumerical MODE). We settled on the waveguide design shown in Fig. \ref{fig:concept}d, which was fabricated at the \emph{Ligentec} foundry. When the seed frequencies enter the waveguide they undergo ND-CFWM, with the separation between lines being determined by the initial separation between the tunable LDs frequencies $\Delta f$ (Fig. \ref{fig:concept}e). As depicted in Fig. \ref{fig:concept}a, the nonlinear generation of comb lines occurs progressively upon propagation through the integrated waveguide.  Given the non-negligible fiber lengths in the DOPA, some initial ND-CFWM occurs before the seed enters the chip, which helps generating a broader comb in the chip but does not alter the waveguide's comb generation mechanism (Supplementary Section I). In the time domain, the coherent sum of the lines of the broadband comb results in a train of ultrashort pulses of pulse duration $\tau''$ $-$ now on the order of hundreds of femtoseconds $-$ with a repetition period $T''=T'=1/ \Delta f$. Our phase-resolved simultaneous spectral and temporal characterization presented below confirms this coherence characteristic. 


\begin{figure}[bp]
    \centering
    \includegraphics[width=0.8\textwidth]{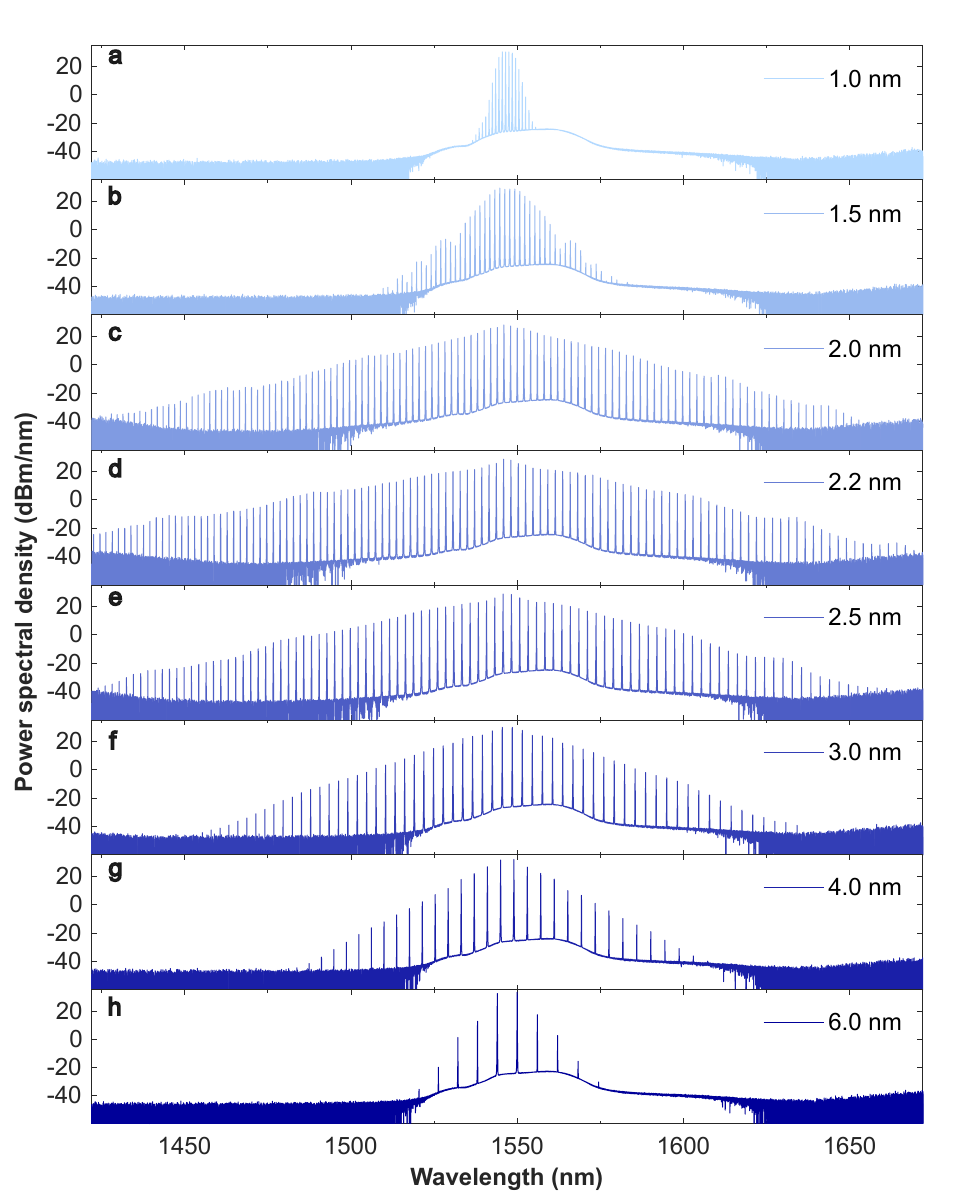}
    \caption{\textbf{Continuous tunability of the line spacing. (a-h)} Measured spectra at the output of the waveguide for pump wavelengths spacing of 1, 1.5, 2, 2.2, 2.5, 3, 4, and 6\,nm center around 1547\,nm.}
    \label{fig:tunability}
\end{figure}

\subsection{Continuously tunable line-spacing}

In the absence of a cavity, the comb line spacing is decoupled from cavity resonant frequencies. 
Instead, it is determined by the frequency separation $\Delta f$ between the two initial tones, enabling continuous tunability over a wide range of detunings. To demonstrate this capability, Fig. \ref{fig:tunability} shows the measured spectra of the frequency combs generated at the output of the chip when the wavelength spacing between the two initial tones is varied from  $\Delta \lambda$ = 1 nm to 6 nm, corresponding to a frequency spacing spanning from 125 GHz to 752 GHZ. The specific values for wavelength spacing within this range $-$ 1, 1.5 nm, 2nm, 2.2 nm, 2.5 nm, 3 nm, 4 nm, and 6 nm, as shown in Fig. \ref{fig:tunability}a-h, respectively $-$ have been chosen to be in different increments to show that the line spacing does not need to be a multiple of an initial FSR. The on-chip coupled power remained constant for all measurements at 88 mW to showcase the impact of the line spacing alone. While continuous line-spacing tunability in microcombs has been shown before \cite{xue2016thermal,joshi2016thermally,he2023high, Shu2023}, the range has been limited to within one FSR, with the widest prior range reported, to the best of our knowledge, being 97.5 GHz \cite{Shu2023}. 
Here we show continuous tunability over a 627 GHz range, only limited by the dispersion of this specific waveguide, which prevents efficient phase-matching for the ND-CFWM process outside that range. 

The results in Fig. \ref{fig:tunability} also showcase the existence of an ideal phase-matching condition: a specific line spacing for which the ND-CFWM process leads to the broadest bandwidth and the largest number of generated lines. This condition is met at 2.2 nm wavelength separation between the two initial tones ($\Delta f = 276$ GHz) in this specific waveguide (Fig. \ref{fig:tunability}d), as dictated by the dispersion characteristics shown in Fig. \ref{fig:concept}b. Efficient cascaded ND-CFWM process is preserved in the range from 2 nm to 3 nm wavelength spacing (Figs. \ref{fig:tunability}c-f). Beyond that range, ND-CFWM still occurs, but its efficiency (at the fixed on-chip coupled power) decreases, as evidenced by the reduced bandwidth of the spectra observed in Figs. \ref{fig:tunability}a,b,g, and h.

To explain this behavior we resort to the expression for the nonlinear parametric gain, $g=\sqrt{4\gamma^2 P_1P_2-(\kappa/2)^2}$, where $\gamma$ is the nonlinear parameter, $P_1$ and $P_2$ are the pump powers at $f_1$ and $f_2$, and $\kappa=\Delta k+\gamma(P_1+P_2)$ is the effective mismatch \cite{agrawal2013}. At fixed pump powers, the maximum parametric gain occurs when $\kappa=0$, which implies that $\Delta k=-\gamma(P_1+P_2)$, where $\Delta k$ accounts for the phase mismatch between the four waves involved in the initial FWM process arising from the dispersion.
In this experiment this condition is met at $\Delta f = 276$ GHz ($\Delta \lambda = 2.2$ nm), which is in qualitative agreement with our nonlinear Schrödinger equation simulations (Supplementary Section II). 


\subsection{High-power broadband comb generation}
\begin{figure}[H]
    \centering
    \includegraphics[width=\textwidth]{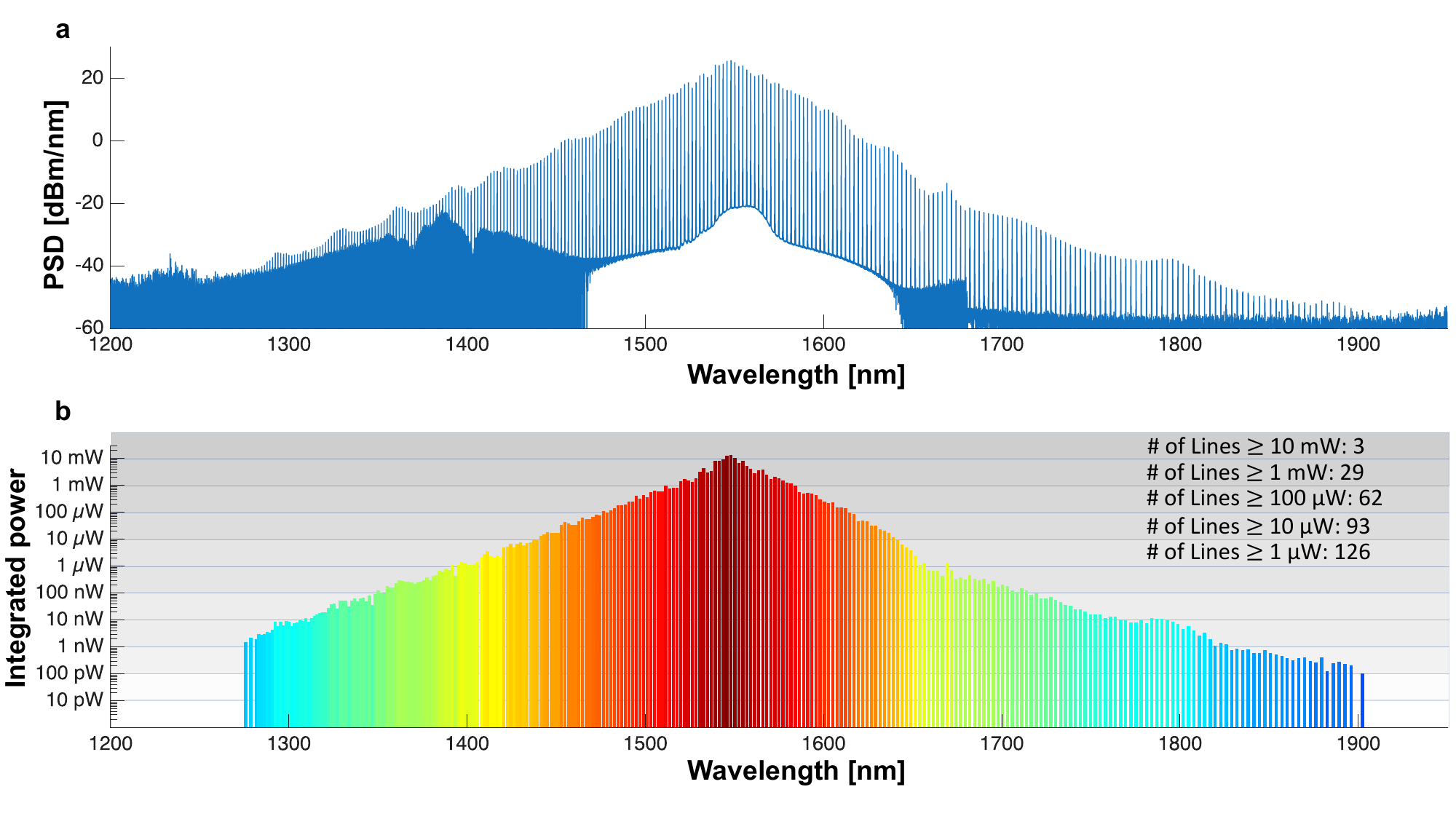}
    \caption{\textbf{High-power per line broadband on-chip comb generation.} \textbf{a}, Power spectral density as a function of wavelength. \textbf{b}, Integrated power per line, excluding background noise. The height and color of each bar represents the power of the line at that spectral position.}
    \label{fig:mega spec}
\end{figure}

Once the optimal line spacing of this experimental platform has been determined, we explore the power dependence of the bandwidth and analyze the power spectral density (PSD) characteristics of the comb. As expected from the power dependence of any nonlinear optical process, increasing the input on-chip coupled power leads to the generation of a larger number of lines and thus a broader bandwidth at a fixed line spacing (Supplementary Section I). Knowing that the on-chip coupled power must be maintained below $\approx 160$ mW to avoid damage in this kind of integrated waveguide, we fixed the DOPA output to just below this level while adjusting its duty cycle -- $\tau/T$ -- and thus its peak power to produce the broadest possible comb at the output of the chip. Specifically, with an on-chip coupled power of 153 mW, $\tau = 30$ ns, $T=10\mu$s, and thus $\tau/T=0.003$, the measured spectrum at the output of the chip spanned from 1280 nm to 1903 nm (a bandwidth of $\approx$623 nm) with 283 lines, as shown in Fig. \ref{fig:mega spec})a. 

To accurately determine the PSD we integrated the power in a narrow spectral window around the center of each line (Supplementary Section III). In this way, we ensure that we obtain the true power contained in each of the lines, not including the power from the background amplified spontaneous emission. The results of this analyisis, shown in Fig. \ref{fig:mega spec}b reveal that sixty-two of the comb lines have more than 100 $\mu$W optical power, twenty-nine lines exceed 1 mW, and three lines exceed 10 mW. This is an improvement of over a factor of two in terms of the number of lines above 100 $\mu$W with respect to the state of the art \cite{gil2025high} while using similar input average power. More crucially, this is the first on-chip comb capable of delivering a significant number of lines above 1 mW. 

The full redistribution of the power in the seeds to the comb, as evidenced by the fact that the two initial lines blend into the comb spectrum, suggests a very high inherent pump-to-comb conversion efficiency. To quantify this efficiency, we resort to the commonly accepted metric $\eta=P_{other} ^{out}/P_{pump} ^{in}$, where $P_{pump} ^{in}$ is the pump power in the input waveguide and $P_{other} ^{out}$ is the power of the other comb lines, excluding the pump lines, at the waveguide output \cite{Xue2017LPR}. Here, $P_{pump} ^{in}=153$ mW and $P_{other} ^{out}=120.85$ mW (calculated directly from the summation of the power of all other lines in Fig. \ref{fig:mega spec}b).  The efficiency of our comb system is, therefore, $\eta=79\%$. For this type of comb, however, given that the two lines coinciding with the two original seed lines are usable $-$ fully integrated in the comb envelope with similar power to the rest of the lines $-$  
it begs the question whether other metrics could also be useful. For instance, we note that from the total power at the output of the chip (150 mW before out-coupling), $98.2\%$ is contained in the comb lines, with only 1.8$\%$ of the chip's output power coming from background amplified spontaneous emission (ASE).



\subsection{Coherence characteristics of the comb}



The time-domain behavior of a frequency comb is governed by the relative phase pattern among its spectral lines and by the stability of that pattern over time. When the comb lines are phase-locked and exhibit a linear phase relationship, the comb produces a stable periodic pulse in the time domain. In a four-wave-mixing-generated comb, the phases of newly generated lines are not random; they are inherited from the driving fields through the nonlinear polarization. For example, the idler generated at $\omega_i = 2\omega_1 - \omega_2$ acquires a phase $\phi_i = 2\phi_1-\phi_2$, while the opposite sideband (signal) at $\omega_s = 2\omega_2-\omega_1$, has a phase $\phi_s = 2\phi_2-\phi_1$. In our system, the two driving fields are independent CW lasers, thus they exhibit an arbitrary relative phase ($\phi_2-\phi_1)$. As the CFWM process generates new lines, the seed phase difference is transferred to each new signal/idler pair in the comb and sets the slope of the linear spectral phase ramp. For equally spaced comb lines indexed such that the two seed lasers are m=1 and m=2, the cascaded comb phase relation is $\phi_m = \phi_1+(m-1)(\phi_2-\phi_1)$. This linear phase ramp corresponds to a temporal shift of the entire pulse train but does not alter the pulse shape. To probe the existence of these dynamics at the output of our chip we performed phase-resolved spectro-temporal measurements using a second-harmonic generation (SHG) frequency-resolved optical gating (FROG) technique \cite{trebino2000frog}.

\begin{figure}[H]
    \centering
    \includegraphics[width=\textwidth]{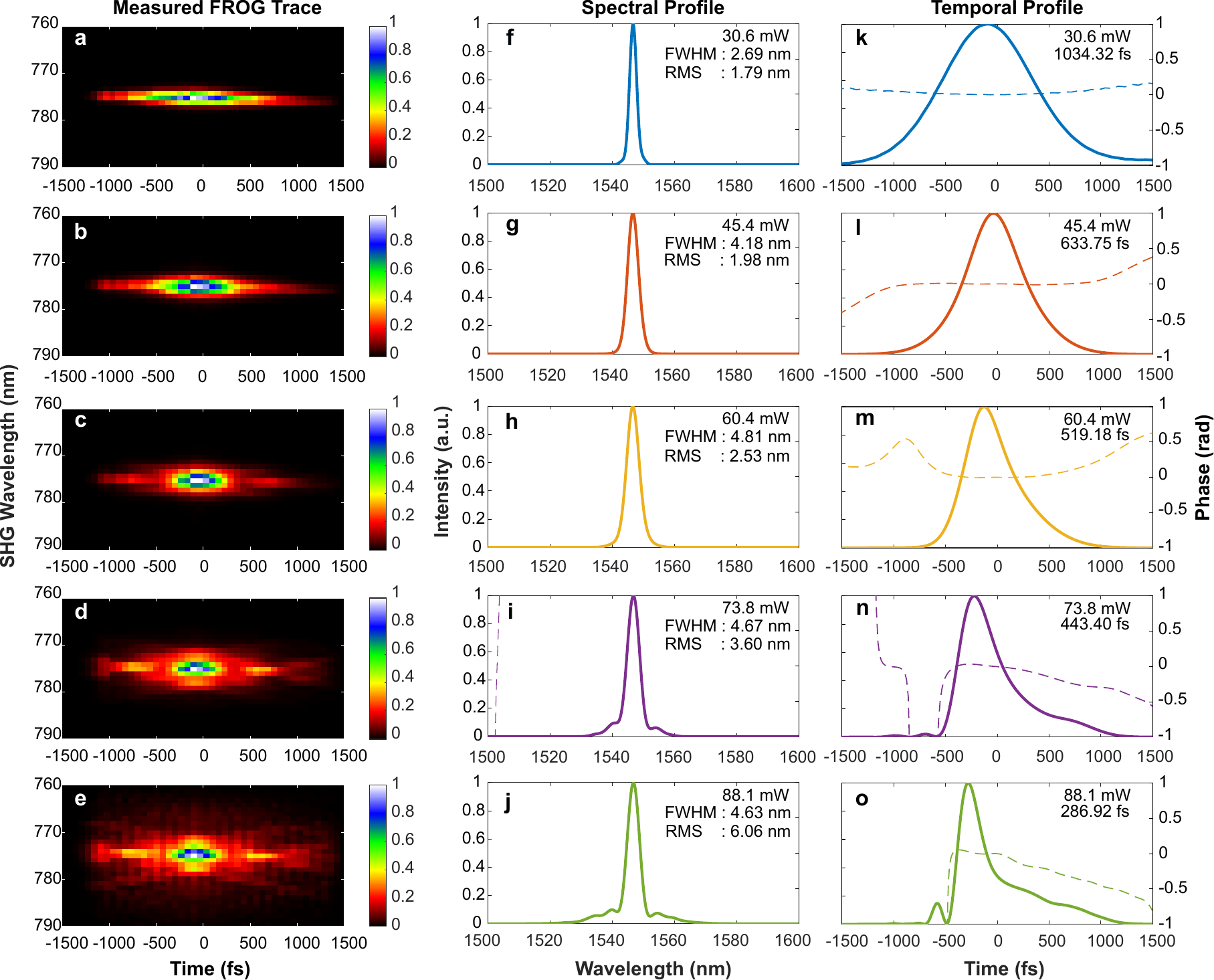}
    \caption{\textbf{Spectro-temporal dynamics of the pulses at the comb output.} \textbf{a-e}, Second-harmonic generation (SHG), frequency resolved optical gating (FROG)-measured spectrograms at five different power levels for a fixed line spacing $\Delta \lambda$=2.2 nm. \textbf{f-j}, Corresponding retrieved spectral intensity of the output pulses. \textbf{k-o}, Corresponding temporal intensity (solid) and phase (dashed) profiles.}
    \label{fig:frog}
\end{figure}

Figures \ref{fig:frog}a-e show the measured SHG-FROG spectrograms (spectrum as a function of delay) at five different power levels at the output of the chip, 30.6, 45.4, 60.4, 73.8 and 88.1 mW, respectively, while keeping a fixed line spacing $\Delta\lambda=2.2$ nm. From the spectrograms we retrieve the spectral intensity (Figs. \ref{fig:frog}f-j) and the temporal intensity and phase profile(Figs. \ref{fig:frog}k-o). In both the spectral and temporal domains, the measurements reveal the existence of clean optical pulses at the output of the chip that are part of a stable pulse train. In the spectral domain we observe that the spectrum of the pulses broadens from 2.69 to 4.81 nm full-width at half maximum (FWHM), as the power increases from low to moderate input powers (Figs. \ref{fig:frog}f-h). This is accompanied by narrowing of the pulses in the time domain from 1034 to 519 fs FWHM (Figs. \ref{fig:frog}k-m). In these three cases, the phase remains flat throughout the duration of the pulse, suggesting nearly transform-limited pulses with time-bandwidth products $\Delta f\Delta\tau$ ranging from 0.34 to 0.313, consistent with the time-bandwidth product of fundamental solitons (0.315). As the power continues to increase, the pulses continue to compress to shorter pulse durations (Figs \ref{fig:frog}n,o), in agreement with the area theorem of solitons, where the pulse energy is inversely proportional to its width $E\propto\tau^{-1}$. Whereas the effects of the third order dispersion (TOD) of the waveguide are negligible for long pulses, at these short pulse durations its effects become noticeable. The peak of the soliton gets advanced in time, developing an asymmetric temporal shape with oscillations toward the leading edge of the pulse as expected from the negative third order dispersion in our waveguide at the center wavelength. At these high powers, the spectrum of the pulses remains symmetric and, while its FWHM decreases slightly, the pulses develop wings and the total spectral content increases, as evidenced by larger root mean square (RMS) values (Figs \ref{fig:frog}i, j).

Overall, the FROG retrievals show clean nearly hyperbolic secant pulses with flat phases across their pulse durations. This indicates that the comb lines within the measured spectrum are mutually phase locked and form a coherent ultrashort pulse. While FROG does not probe long-term phase noise or the absolute stability of individual comb teeth, the retrieved phase demonstrates high mutual coherence of the comb modes contributing to the formation of the train of ultrashort pulses.

\section{Discussion and conclusions}\label{sec13}
We demonstrated a cavity-less approach for optical comb generation in dispersion-engineered integrated photonic waveguides capable of delivering broadband coherent-state combs with high pump-to-comb conversion efficiency and power per line. While high-Q microresonator combs have transformed integrated frequency-comb generation and are excellent for low-threshold applications, the approach presented here addresses a bottleneck that becomes increasingly important as combs move from metrology demonstrations to deployed photonic systems: per-channel optical power. In high-parallelism AI accelerators, optical interconnects, RF photonics, and coherent optical communications, each comb tooth must carry information, drive a device, or generate a detectable signal. Low per-line power forces additional amplification, increases noise and power consumption, and reduces the effective precision or link margin.

A key element of this initial demonstration has been the use of a DOPA seed to deliver high peak powers while maintaining moderately low average powers and linewidth. In the experiments presented here, the DOPA was implemented using off-the-shelf tunable laser diodes, modulators, and fiber components. All these components are, however, readily available in their integrated photonics form: III-V/silicon tunable lasers can deliver $>$20 mW with tuning ranges well above C-band \cite{su2025heterogeneously}; high-extinction carrier-based modulators are available in mainstream silicon photonics foundries \cite{uddin2023edited} and Pockels-based modulators are slowly becoming parts of the foundries offering too \cite{xu2020high, hou2024high}; finally, on-chip amplification is available either via semiconductor optical amplifiers \cite{davenport2016heterogeneous, van201927} or directly in the silicon nitride layer via rare-earth dopants \cite{liu2022photonic, che2023first_edited}. There is thus a clear path towards a fully on-chip version of the cavity-less comb approach presented here, much like microcomb research evolved from initial demonstrations where only the microresonator was on-chip \cite{del2007optical} to the current on-chip system demonstrations  \cite{gil2025high, ling2024electrically}.  

Future research includes further optimization of the waveguide dispersion to meet desired metrics. For instance, tuning the dispersion to achieve optimal phase-matching at line-spacings matching the dense wavelength division multiplexing (DWDM) standard spacing 100 GHz, octave-spanning bandwidths, or flatter comb spectra. An exhaustive exploration of the parameter space of seed laser duty cycle, pulse duration, and repetition rates is expected to open avenues in the optimization of other metrics such as power per line and linewidth.   

\backmatter
\bmhead{Acknowledgments} Research was sponsored by the Army
Research Office and was accomplished under Grant Number W911NF-26-1-A162. The views and
conclusions contained in this document are those of the authors and should not be interpreted as
representing the official policies, either expressed or implied, of the Army Research Office or the U.
S. Government. The U.S. Government is authorized to reproduce and distribute reprints for
Government purposes notwithstanding any copyright notation herein. The authors also acknowledge support from the Air Force Research Laboratory (FA86511820019).



\bibliography{sn-bibliography}

\end{document}